\documentclass[journal]{IEEEtran}
\ifCLASSINFOpdf
   \usepackage[pdftex]{graphicx}

\else

\fi

\usepackage{xcolor}
\usepackage{mathtools}
\usepackage{cuted}
\usepackage{booktabs}
\usepackage{amsfonts}
\usepackage{amssymb}
\usepackage{amsmath, bm}
\usepackage{cite}
\usepackage{amssymb, amsmath, amsthm, amsfonts}
\usepackage{caption}
\usepackage{subcaption}
\newcommand{\bx}{\mathbf{x}}
\newcommand{\bz}{\mathbf{z}}
\newcommand{\btheta}{\boldsymbol{\theta}}
\newcommand{\bmu}{\boldsymbol{\mu}}
\newcommand{\bV}{\mathbf{V}}
\newcommand{\bP}{\mathbf{P}}
\newcommand{\bQ}{\mathbf{Q}}

\newcommand{\st}{\text{s.t.}}

\begin{document}
\title{Learning to Solve the AC Optimal Power Flow via a Lagrangian Approach}

\author{
\IEEEauthorblockN{Ling Zhang, Baosen Zhang}
\IEEEauthorblockA{ Department of Electrical and
Computer Engineering \\
University of Washington, Seattle WA, USA\\
\{lzhang18, zhangbao\}@uw.edu}
}

\maketitle

\begin{abstract}
Using deep neural networks to predict the solutions of AC optimal power flow (ACOPF) problems has been an active direction of research. 
However, because the ACOPF is nonconvex, it is difficult to construct a good data set that contains mostly globally optimal solutions.
To overcome the challenge that the training data may contain suboptimal solutions, we propose a Lagrangian based approach. First, we use a neural network to learn dual variables of the ACOPF problem. Then we use a second neural network to predict solutions of the partial Lagrangian from the predicted dual variables. Since the partial Lagrangian has a much better optimization landscape, we use the predicted solutions from the neural network as a warm start for the ACOPF problem. Using standard and modified IEEE 22-bus, 39-bus and 118-bus networks, we show that our approach is able to obtain the globally optimal cost even when the training data is mostly  comprised of suboptimal solution.

\end{abstract}



\section{Introduction}
The AC optimal power flow (ACOPF) problems are fundamental to power system operations, but they are often computationally expensive to solve in real-time~\cite{castillo2013computational}. 
Recently, machine learning has emerged as a popular method to aid in solving ACOPFs~\cite{pan2020deepopf,guha2019machine,owerko2020optimal,zhou2020data}. By treating the optimization problem as a function that maps loads to the optimal solutions (voltage and angles), supervised learning techniques can be used to train a neural network (NN) that replaces a nonlinear programming solver~\cite{singh2021learning,baker2020emulating,lange2020learning,zamzam2019learning}. Since training can be done offline using historical or simulated data, neural networks can be used in real-time to reduce computational burdens. 

Because the solution of the ACOPF problem need to satisfy a set of nonlinear constraints, neural networks typically cannot guarantee that their outputs are feasible. Therefore, most needs an additional power flow or OPF step to obtain feasible solutions~\cite{baker2020emulating,donti2021dc3}. Equivalently, we can think of these machine learning methods to be producing a good \emph{warm start} for a solver. Since the starting point is of critical significance to a nonconvex problem, a good initialization would offer significant computational speedups in real-time~\cite{bertsekas1997nonlinear}. 

A key, and often understated assumption, in using learning is that the training set if of high quality. In the context of ACOPF, the training set consists of pairs of active/reactive loads and their corresponding ACOPF solutions, and the assumption is that the solution are the optimal ones. However, the training set is typically constructed using existing noninear programming solvers, and there is no guarantee that the solutions are in fact globally optimal. Because ACOPF is not convex, there could exist multiple local solutions. These solutions can lead to large differences in the cost function, and there is no tractable way to judge the quality of solutions for a given problem. Therefore, the quality of the learned solutions are fundamentally limited by the quality of the training data sets. 

In addition to the presence of suboptimal solutions, a more challenging setting for learning would be to have several solutions associated with loads that are close to each other. Because of the presence of multiple solutions, a small change in load may lead a solver to jump between distinct solutions. Therefore, the training set could include data that are close in the load, but quite different in the solutions. For a neural network trained using regression loss, it would output the average of the local solutions, leading to a initialization point that neither increase the computation speed or help with the quality of the solutions. 

In this paper, we present a machine learning architecture that overcomes the challenge of multiple suboptimal local solutions in the training data set. Instead of focusing on the load/solution pairs, we seek to learn a neural network that maps load to the dual variable of the power balance constraints. These dual variables are the locational marginal prices, and would be readily available from any modern nonlinear solver. These dual variables are used to form a partial Lagrangian, whose solution we also learn via a neural network. All together, we use two neural networks, one mapping the load to the dual variables, another mapping the predicted dual variables and load to the solution of a partial Lagrangian. 

We use the predicted solution of the partial Lagrangian as a warm start. Interestingly, this warm starting point tends to be closer to the globally optimal solution of the ACOPF, even if the training data set only has suboptimal solutions or a mixture of global and local solutions. Intuitively speaking, the optimization landscape of the partial Lagrangian is much ``easier'' than the original ACOPF problem, and its solution is robust to errors in the Lagrangian multiplier. Therefore, the quality of the learned warm start can also be potentially much better than directly learning based on load/solution pairs. 

In this paper, we show that our duality-based approach outperforms existing approaches on modified IEEE 22, 39 and 118 buses networks~\cite{bukhsh2013local}. The difference is especially significant if the training data set contains some strictly suboptimal solutions. Our method maintains the advantage of using neural networks to emulate solvers: it provides a good warm start in real-time. It also has the added benefit of improving the solution quality in the training data set. Our paper is organized as follows. Section~\ref{sec:model} states the problem definition, Section~\ref{sec:algo} gives the main algorithm, Section~\ref{sec:geometry} provides some geometric intuition behind the algorithm and Section~\ref{sec:results} shows the simulation results.

\section{Problem Formulation} \label{sec:model}
\subsection{ACOPF Formulation}
Consider a power system network where $n$ buses are connected by $m$ edges. 
For bus $i$, let $V_i$ denote its voltage magnitude, $\theta_i$ its angle, $P^{G}_i$ and $Q^{G}_i$ the active and reactive output of the generator and $P^{D}_i$ and $Q^{D}_i$ the active and reactive load. We use $P^{f}_{ij}$ and $Q^{f}_{ij}$ to denote the active and reactive power flowing from bus $i$ to bus $j$. The admittance between buses $i$ and $j$ is $g_{ij}-b_{ij}$. 
We use $\theta_{ij}$ as a shorthand for $\theta_i-\theta_j$. 

The ACOPF problem is to minimize the cost of active power generations while satisfying a set of constraints~\cite{bukhsh2013local}:
\begin{subequations} \label{prob1}
\begin{align}
     \min_{\bV, \btheta} &\textstyle \sum_{i} c_i(P^{G}_i)\\
    \st ~ & P^{G}_i = P^{D}_i + \textstyle \sum_{j=1}^{N} P^{f}_{ij}\label{Pbalanc}\\
    & Q^{G}_i = Q^{D}_i + \textstyle \sum_{j=1}^{N} Q^{f}_{ij}\label{Qbalanc}\\
    & P^{f}_{ij} = V_i^2g_{ij}-V_iV_j(g_{ij}\cos(\theta_{ij})-b_{ij}\sin(\theta_{ij}))\label{PEq}\\
    & Q^{f}_{ij} = V_i^2 \hat{b}_{ij} - V_iV_j(b_{ij}\cos(\theta_{ij}) +g_{ij}\sin(\theta_{ij}))\label{QEq}\\
    & \underbar{V}_i\leq V_i \leq \bar{V}_i\label{Vlimits}\\
    & \underbar{P}^{G}_i\leq P^G_i \leq \bar{P}^{G}_i\label{PGlimits}\\
    & \underbar{Q}^{G}_i\leq Q^G_i \leq\bar{Q}^{G}_i\label{QGlimits}\\
    & (P^{f}_{ij})^2+(Q^{f}_{ij})^2\leq (S_{ij}^{\max})^2\label{flimits}
\end{align}
\end{subequations}
where $\hat{b}_{ij}=b_{ij}-0.5b_{ij}^C$ and $b_{ij}^C$ is the line charging susceptance. The constraints  (\ref{Pbalanc}) and (\ref{Qbalanc}) enforce power balance, (\ref{PEq}) and (\ref{QEq}) are the AC power flow equations, (\ref{Vlimits}) limits the bus voltage magnitudes, (\ref{PGlimits}) and (\ref{QGlimits}) represent the active and reactive limits and \eqref{flimits} are the line flow limits. 

\subsection{Learning Neural Networks}
It is useful to view the optimization problem in \eqref{prob1} as a mapping from the demands $P^D$ and $Q^D$ to the solutions $\bV$ and $\btheta$. Machine learning is typically used to find a proxy of this mapping by training (deep) neural networks. Suppose \eqref{prob1} is solved for a number of demands and the corresponding solutions are collected into a training data set. Several approaches exist for training, including direct regression~\cite{pan2020deepopf}, the emulation of a iterative solver~\cite{baker2020emulating}, and using sensitivity information~\cite{singh2021learning}. The constraints in \eqref{Vlimits}, \eqref{PGlimits} and \eqref{QGlimits} can be satisfied by using $\tanh$ or sigmoid activation functions in the output layer of the neural network. But the rest of the constraints would need to be satisfied by calling another power flow or optimal power flow step~\cite{baker2020emulating,donti2021dc3}.

However, all of the above methods face a common challenge, stemming from the fact that the ACOPF problem in~\eqref{prob1} is nonconvex and may have multiple solutions~\cite{bukhsh2013local,wu2017deterministic,Molzahn19}. A common assumption has been that during normal operations, the ACOPF would have only one solution with ``practical'' values. However, multiple recent works have shown that for reasonable operation conditions, there could be multiple solutions that differ significantly in cost, but all have practical values (e.g., with voltages being all close to 1 p.u.)~\cite{ma1993efficient,momoh1999review,lesieutre2015efficient,lesieutre2019distribution}.

Assuming \eqref{prob1} is feasible, it can have two class of solutions: local solutions and global solutions. Local solutions are all the solutions that satisfies local optimally conditions, for example, the KKT conditions or second order ones~\cite{bertsekas1997nonlinear}. Out of this set, the solutions with the lowest cost are called the global ones. We sometimes refer to the local solutions that are not global as strict local solutions. 

Over the years, many nonlinear programming (NLP) solvers have been developed for the ACOPF problem, and their speed and efficiency have improved dramatically (e.g.,~see~\cite{cain2012history} and the references within). However, NLP solvers are typically only able to return local solutions and there is generally no way to tell whether they are globally optimal or not. Therefore, a training data set created using NLP solvers may fundamentally limit the solution quality when a neural network is used. Importantly, the solution returned by a NLP solver is sensitive to a variety of factors, and small changes in demand can lead to a large change in the solution. For example, small changes in the demand can lead to the solution switching between global and local. Therefore, a data set may very well consist of a mixture of local and global solutions, which tend to be very confusing for the neural network to learn. 




\section{Algorithm} \label{sec:algo}
In this section, we describe our learning approach to find more optimal solutions to ACOPF problems using neural networks, even when the training data set contains a mixture of local and global solutions. We first give a partial Lagrangian-based approach, then we discuss the learning model and the training of neural networks. We also show how to obtain the optimal solution to the ACOPF problem using the trained neural networks.

\subsection{Lagrangian-based Approach}
In paper \cite{zhang2021iterative}, we propose an iterative approach to improve the solution quality by alternatively solving (\ref{prob1}) and its partial Lagrangian. 
The partial Lagrangian for (\ref{prob1}) is formed by dualizing the active and reactive power balance constraints \eqref{Pbalanc} and \eqref{Qbalanc}. Suppose the Lagrangian multipliers associated with \eqref{Pbalanc} and \eqref{Qbalanc} are $\mu^{P}_{i}$ and $\mu^{Q}_{i}$, respectively, then the partial Lagrangian for \eqref{prob1} is:
\begin{subequations}\label{eq:Lagrangian}
\begin{align}
     \mathcal{L}_{\bmu} (\bV, \btheta) = &\textstyle \sum_{i} c_i(P^{G}_i)+\sum_{i}\mu^{P}_{i}(P^{D}_i + \textstyle P^{f}_{i}-P^{G}_i)\nonumber\\
     & + \sum_{i}\mu^{Q}_{i}(Q^{D}_i + \textstyle Q^{f}_{i}-Q^{G}_i)\\
         \st ~ & \underbar{V}_i\leq V_i \leq \bar{V}_i\\
    & \underbar{P}^{G}_i\leq P^G_i \leq \bar{P}^{G}_i\\
    & \underbar{Q}^{G}_i\leq Q^G_i \leq\bar{Q}^{G}_i\\
    & (P^{f}_{ij})^2+(Q^{f}_{ij})^2\leq (S_{ij}^{\max})^2
\end{align}
\end{subequations}
where $P^{f}_{i}=\sum_{j=1}^{N} V_i^2g_{ij}-V_iV_j(g_{ij}\cos(\theta_{ij})-b_{ij}\sin(\theta_{ij})) $, and $Q^{f}_{i} = \sum_{j=1}^{N} V_i^2 \hat{b}_{ij} - V_iV_j(b_{ij}\cos(\theta_{ij}) +g_{ij}\sin(\theta_{ij}))$ are the AC power flow equations.

The critical insight in~\cite{zhang2021iterative} is that the solution of the partial Lagrangian in~\eqref{eq:Lagrangian} tend to be close to the global optimal solution of the original ACOPF in~\eqref{prob1}. This is true even if the multipliers $\mu^P$ and $\mu^Q$ are the ones associated with the strict local solutions (see~\cite{zhang2021iterative} for more details). Therefore, the solution of \eqref{eq:Lagrangian} is a good warm start point for an ACOPF solver. In~\cite{zhang2021iterative}, an iterative algorithm was proposed, where an ACOPF and its partial Lagrangian are repeatedly and iteratively solved to move from local solutions to global ones. 

In this paper, we use learned neural networks to replace explicitly solving (\ref{prob1}) and \eqref{eq:Lagrangian}. Namely, we train two neural networks, the first to predict the dual variables of the active and reactive power balance constraints from the load, and the second to predict the solution of \eqref{eq:Lagrangian} from the load and the predicted multipliers from the first neural network. The output of the second neural network is used as a warm start point for an ACOPF solver. Since this starting point is close to the optimal solution, the ACOPF solver is solved much faster and may return a more optimal solution than a solver with a flat or random start. 


\subsection{Training of Neural Networks}
\begin{figure}
\centering
\begin{subfigure}[b]{0.3\textwidth}
 \centering
 \includegraphics[height=3cm, width=5cm]{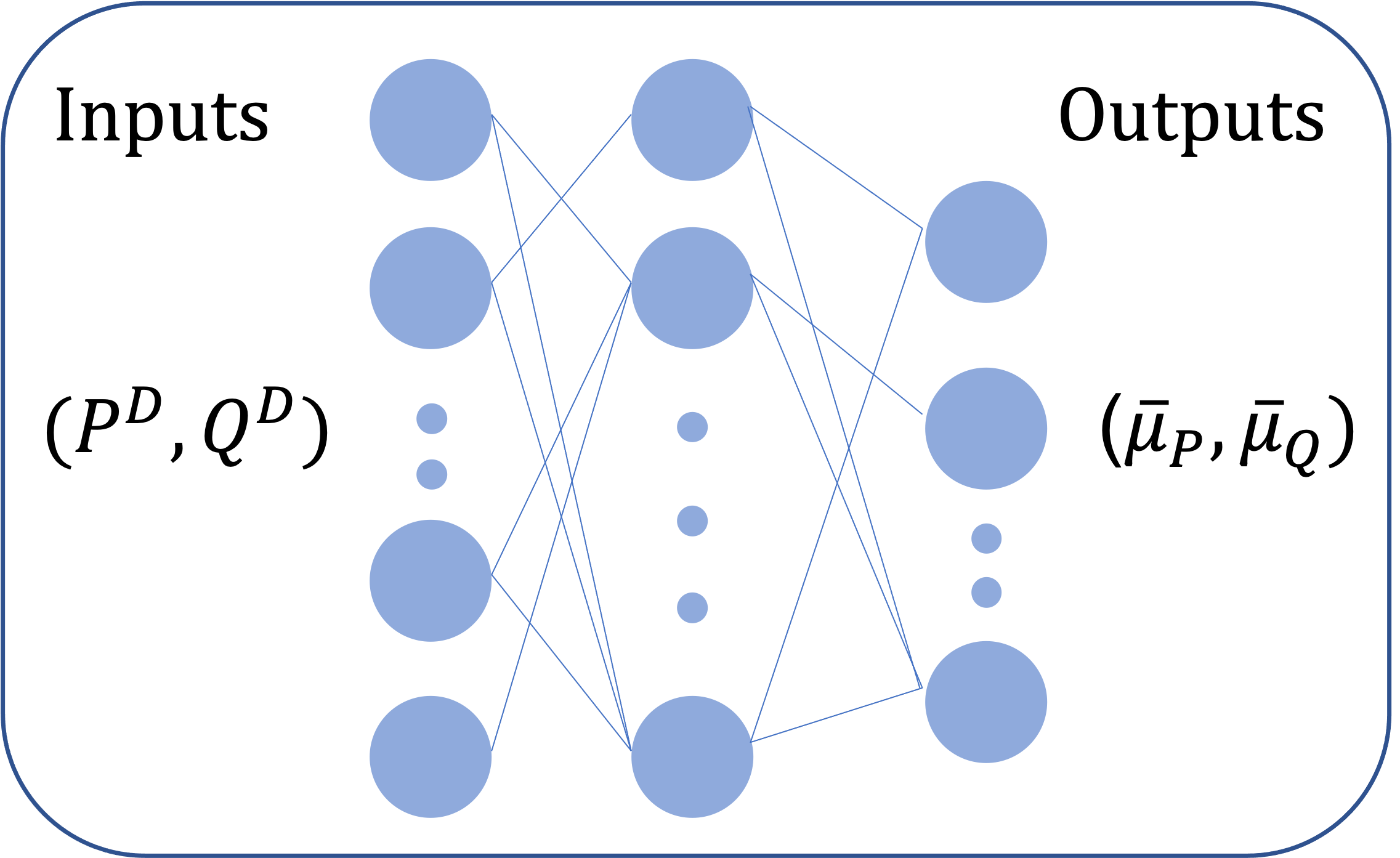}
 \caption{The learning model to emulate solutions to (\ref{prob1}).}
 \label{fig:neural network1}
\end{subfigure}
\hfill
\begin{subfigure}[b]{0.3\textwidth}
 \centering
 \includegraphics[height=3cm, width=5cm]{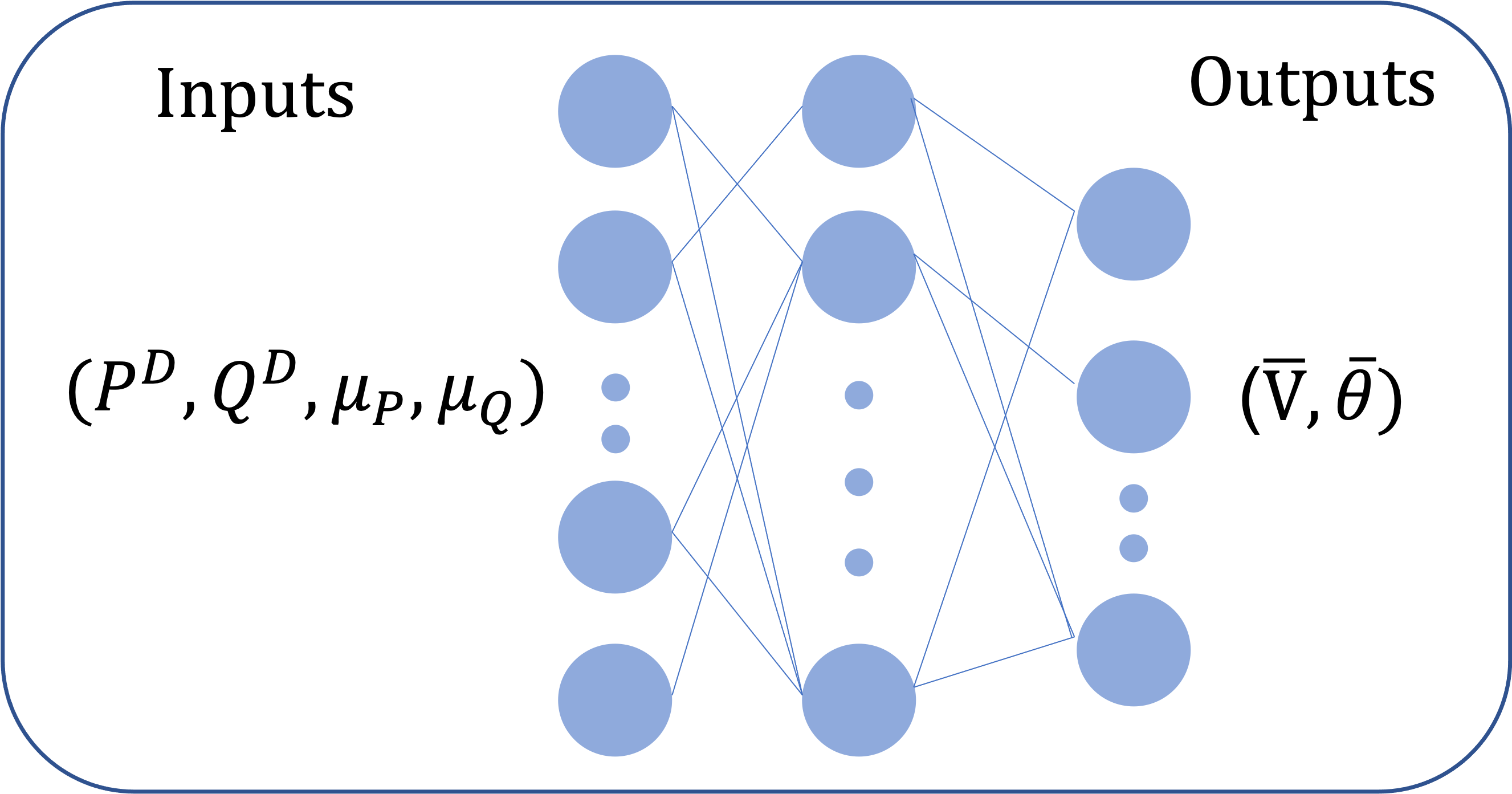}
\caption{The learning model to emulate solutions to the partial Lagrangian \eqref{eq:Lagrangian}.}
\label{fig:neural network2}
\end{subfigure}
\caption{The two neural networks to be trained.}
\vspace{-1.5em}
\label{fig:training}
\end{figure}
Our algorithm treats (\ref{prob1}) and its partial Lagrangian \eqref{eq:Lagrangian} as two operators and train two neural networks separately to emulate them and provide the solutions. The two models of neural network are shown in Fig.~\ref{fig:training}.
For the first neural network to approximate the solution of (\ref{prob1}), the input is the active and reactive load demands $(\bm{P}^{D}, \bm{Q}^{D})$, and the output is the predicted multipliers, denoted by $(\bar{\bmu}_{P}, \bar{\bmu}_{Q})$. Let $\bx$ be the collection of voltage magnitudes and angles, $\bmu$ be the collection of $(\bmu_{P}, \bmu_{Q})$, and $({\bx}^{i}, {\bmu}^{i})$ be the i-th pair of data in the training set, then the first neural network is trained by minimizing the following mean squared loss:
\begin{align}
    \min_{\bm{w}} \frac{1}{N}\sum_{i=1}^{N}( {\bmu}^{i}-g_{acopf}({\bx}^{i};\bm{w}))^2,
\end{align}
where $g_{acopf}(\cdot)$ represents the neural network mapping the load demand to the dual variables, and $\bm{w}$ denotes all trainable parameters.

Then we train a second neural network to emulate the solutions to the partial Lagrangian \eqref{eq:Lagrangian}.
The input for the second neural network is the active and reactive load demands $(\bm{P}^{D}, \bm{Q}^{D})$, as well as the associated dual variable solutions $(\bmu_{P}, \bmu_{Q})$ to (\ref{prob1}). The output of the neural network is the solution to \eqref{eq:Lagrangian}, denoted by $(\bar{\bV}, \bar{\btheta})$. Let $\bz$ be the collection of inputs, $\bar{\bx}$ be the collection of outputs, and $({\bz}^{i},\bar{\bx}^{i})$ be the i-th pair of data in the training set, then the second network is also trained by minimizing the mean squared loss:
\begin{align}
    \min_{\bm{w}} \frac{1}{N}\sum_{i=1}^{N}( \bar{\bx}^{i}-g_{dual}({\bz}^{i};\bm{w}))^2,
\end{align}
where $g_{dual}(\cdot)$ represents the neural network mapping the load demand and dual variable solutions to the minimizer of \eqref{eq:Lagrangian}.

\subsection{Making Predictions in Real-time}
\begin{figure}[t]
    \centering
    \includegraphics[height=2cm, width=8.5cm]{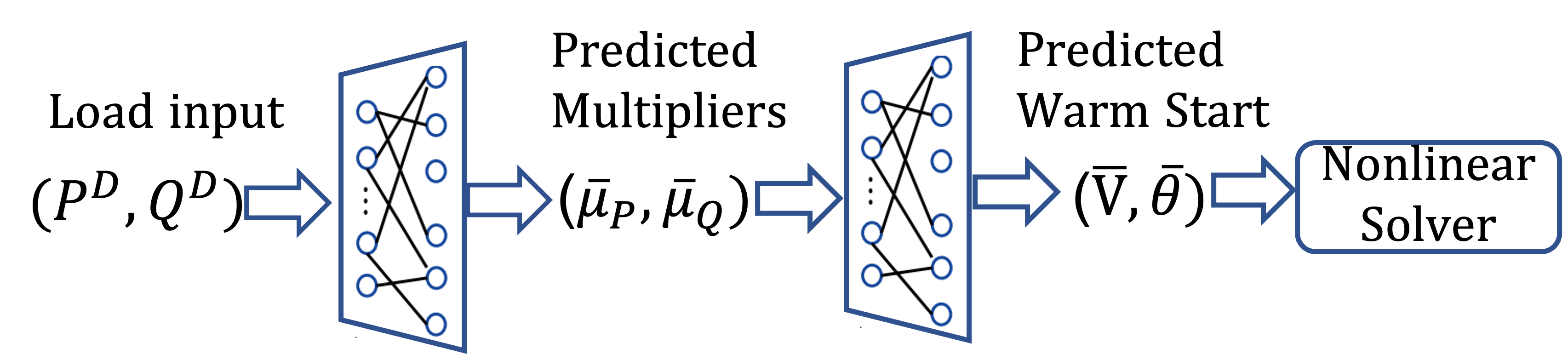}
    \caption{Outline of the solution process.}
    \label{fig:intro}
    \vspace{-0.5cm}
\end{figure}

After training the two neural networks, we use the process in Fig.~\ref{fig:intro} to make predictions in real-time. We use the first trained neural network to predict dual variable solutions $(\bar{\bmu}_{P}, \bar{\bmu}_{Q})$ from the load demand. Then from the predicted dual variable solutions, we use the second trained neural network to predict the solutions $(\bar{\bV}, \bar{\btheta})$ of the Lagrangian. In the end, we call the NLP solver to solve (\ref{prob1}) using $(\bar{\bV}, \bar{\btheta})$ as the initialization. This learning algorithm is summarized below as Algorithm 1. 

\begin{table}[ht]
\normalsize
\begin{tabular}{l}
\hline
\textbf{Algorithm 1:  Solving ACOPF using learning}\\
\hline
\textbf{Inputs:}~$(\bP^{D}, \bQ^{D})$\\
Use the first trained neural network to predict multipliers:\\
\qquad $g_{acopf}({\bP}^{D}, {\bQ}^{D};\bm{w})\longrightarrow (\bar{\bmu}_{P}, \bar{\bmu}_{Q})$\\
Use the second trained neural network to predict solutions \\
to \eqref{eq:Lagrangian}:\\
\qquad $g_{dual}({\bP}^{D}, {\bQ}^{D}, \bar{\bmu}_{P}, \bar{\bmu}_{Q} ;\bm{w})\longrightarrow (\bar{\bV}, \bar{\btheta})$\\
Call NLP solver for \eqref{prob1} initialized at $(\bar{\bV}, \bar{\btheta})$;\\
\textbf{Outputs:}~ Solutions $(\hat{\bV}, \hat{\btheta})$ to \eqref{prob1}.\\
\hline
\end{tabular}
\label{algo1}
\vspace{-0.2cm}
\end{table}

In our algorithm, a data set that must contain a large number of globally optimal solutions is not necessary, since we are predicting high-quality warm starts instead of the ACOPF solutions.
Even when the training data set contains suboptimal solutions, the solution of the Lagrangian would be a good warm start that makes the NLP solver get around being trapped at strictly local solutions. The reason is that the partial Lagrangian in \eqref{eq:Lagrangian} has ``nice'' geometric properties \cite{zhang2021iterative}:
It has minimums that are near the globally optimal solution of the ACOPF problem, even for multipliers obtained at local optimal solutions of the ACOPF problem. 

Also, our algorithm is robust in the sense that the solutions $\bar{\bx}$ to partial Lagrangian are not sensitive to the variations in $\bar{\bmu}$. That is, 
even if the multipliers $\bar{\bmu}$ associated with different local solutions are different, the resulting solutions to partial Lagrangian do not change much. This also means we do not ask the predictions of the neural network to be rather accurate. Therefore, the training set for our algorithm is not necessarily to be very large. We sketch the geometric intuitions behind our algorithm in Section~\ref{sec:geometry}. We validate our algorithm in standard and modified IEEE benchmark systems, and report simulation results in Section~\ref{sec:results}.

\section{Geometry and Intuition}
\label{sec:geometry}

In this section, we use a 2-bus network as an example to shed some light on why Algorithm 1 might learn more globally optimal solutions, even when the training data consists of a lot of local solutions.
The main reason is that Algorithm 1 predicts the solutions of the partial Lagrangian, which would be close to the global minimum of the ACOPF problem regardless of the quality of the training data.
In comparison, a training set with a lot of local solutions would be harmful to the learning process of direct regression method.

In the considered 2-bus network, we ignore the reactive power and set both voltage magnitudes to 1 p.u. for simplicity. 
Suppose bus 1 is a generator and the reference (slack) bus with an increasing cost function $c(\cdot)$, and bus 2 is the load bus with angle $-\theta$. The line admittance is $g-jb$. 
Given a load of $l$ at bus 2 and ignoring all constraints except for the load balancing one, the ACOPF in \eqref{prob1} becomes
\begin{subequations} \label{eqn:two_bus}
\begin{align}
    \min_{\theta} \; & c(g-g\cos(\theta)+b\sin(\theta))\\
    \st~& l+g-g\cos(\theta)-b\sin(\theta)=0\label{eq:2bus}. 
\end{align}
\end{subequations}
This is an example of an OPF with a disconnected feasible space, since there are two discrete solutions to \eqref{eq:2bus} and we are asking for the lower cost one. 

\begin{figure}[ht]
\centering
\includegraphics[height=5cm, width=6cm]{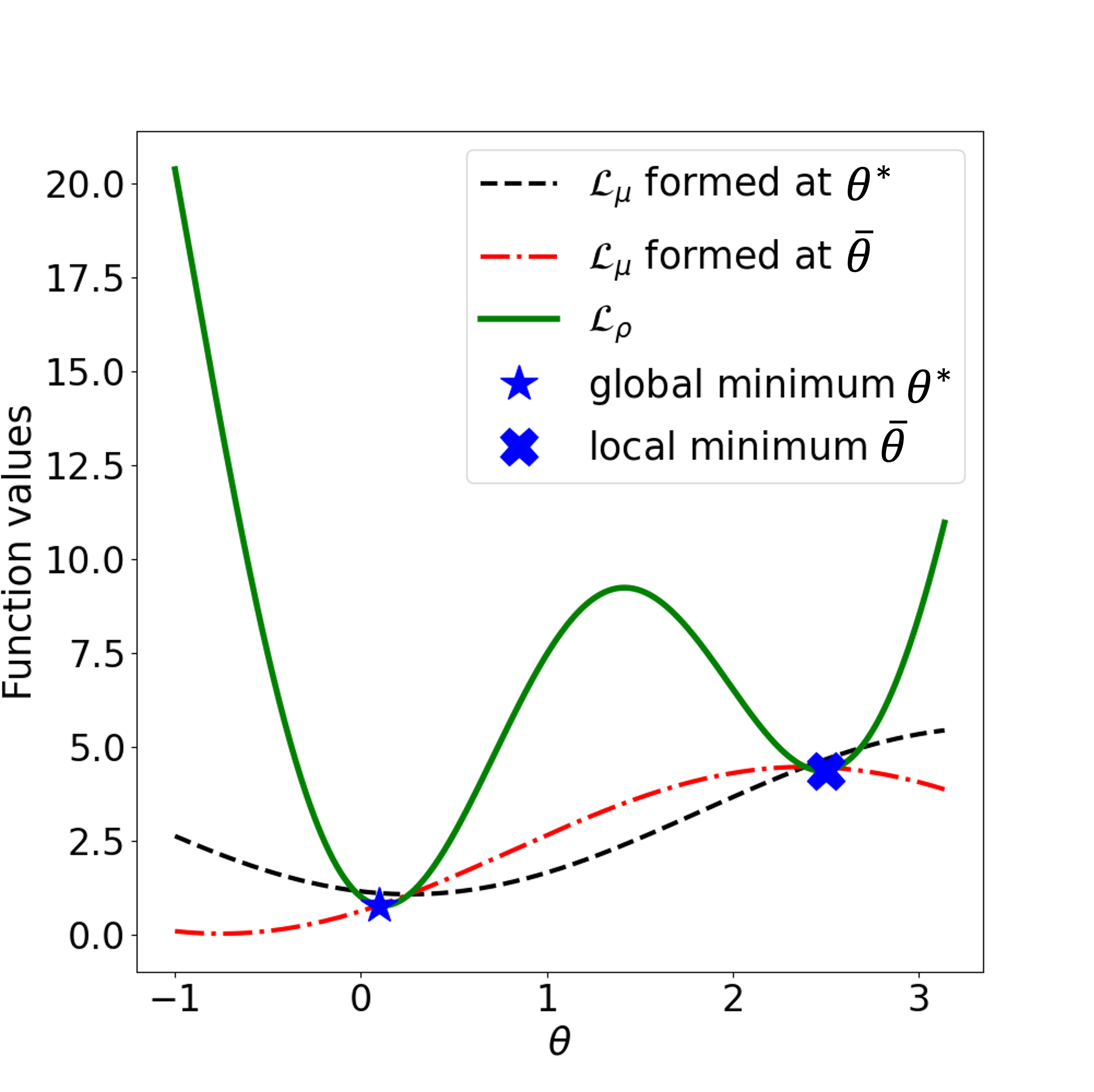}
\caption{Geometry of the penalized objective function $\mathcal{L}_{\rho}$ and the partial Lagrangian $\mathcal{L}_{\mu}$. The line admittance is $g-jb$ and the penalty parameter is $2$. The red curve is the partial Lagrangian formed at the strict local solution and the black curve is at the global solution.
}
\label{fig:geometry}
 \vspace{-0.3cm}
\end{figure}

To see how a NLP solver would approach this problem, we adopt the common practice in nonlinear programming and form a penalized version of~\eqref{eqn:two_bus}~\cite{bertsekas1997nonlinear,mulvaney2020load}. The penalized unconstrained problem is given by 
\begin{align}\label{eqn:pentwobus}
    \mathcal{L}_{\rho}= & c(g-g\cos(\theta)+b\sin(\theta)) \\ 
    &+\rho/2(l+g-g\cos(\theta)-b\sin(\theta))^2, \nonumber
\end{align}
where $\rho$ is a penalty parameter. For large enough $\rho$, the solutions of \eqref{eqn:pentwobus} would coincide with those of~\eqref{eqn:two_bus}~\cite{bertsekas1997nonlinear}. 
The function $\mathcal{L}_{\rho}$ is plotted in Fig.~\ref{fig:geometry}. We can see that there are two local minimas, with the left one being global. 
The strict local minimum (the right one) satisfies both first and second order optimality conditions. Therefore, if we initialize a NLP solver with a poor starting point, it would be stuck at the strict local solution. 

\captionsetup[figure]{font=small,skip=2pt}

Now suppose $\mu$ is the multiplier corresponding to the equality constraint (\ref{eq:2bus}) at \emph{the strict local solution}. The partial Lagrangian of \eqref{eqn:two_bus} by dualizing \eqref{eq:2bus} is:
\begin{align} \label{eqn:Lagrangian2bus}
    \mathcal{L}_{\mu}= & c(g-g\cos(\theta)+b\sin(\theta)) \\
    & +\mu(l+g-g\cos(\theta)-b\sin(\theta)). \nonumber
\end{align}
Since the sinusoidal functions are periodic with period $2\pi$, let us consider the range $\theta\in[-\pi, \pi]$.
It is interesting now to compare the solution of $\mathcal{L}_{\mu}$ and the original problem in~\eqref{eqn:two_bus} (or equivalently, $\mathcal{L}_\rho$). 
The red curve in Fig.~\ref{fig:geometry}
plots $\mathcal{L}_{\mu}$ at the local minimum and the black curve at the global minimum. We can observe an interesting fact that  the minimum of $\mathcal{L}_\mu$ is close to the global minimum of $\mathcal{L}_\rho$, even when the multiplier at the strict local solution is used.

\begin{figure}[ht]
\centering
\includegraphics[height=5cm, width=6cm]{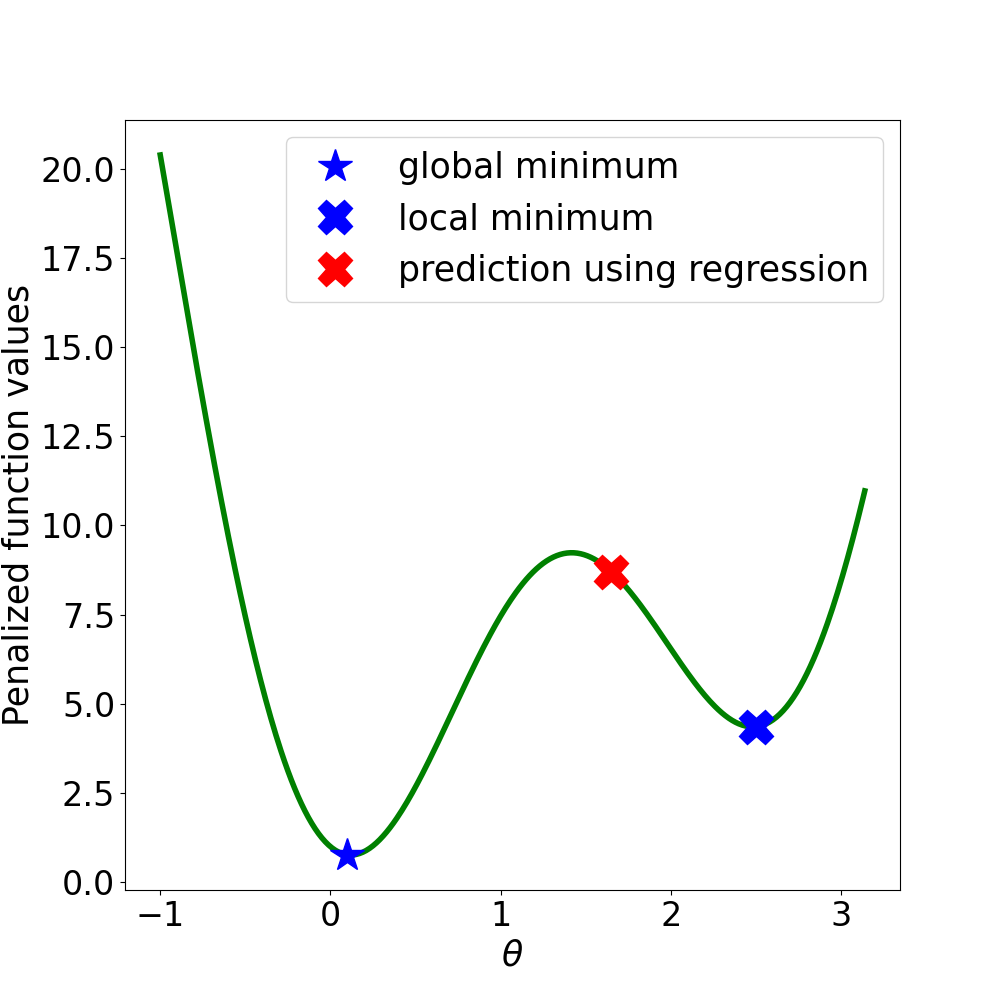}
\caption{Use direct regression to learn a warm start point for the ACOPF solver. When the training set has both local and global solutions, the predicted warm start using direct regression is marked as red, which is close to the strict local minimum of the ACOPF problem. A solver initialized with this predicted warm start would converge to the local solution.}
\label{fig:regression}
\vspace{-0.2cm}
\end{figure}

Let us construct a training data set with both local and global solutions for the same load, and compare the learned warm starts using direct regression and Algorithm 1. Suppose a regression method is used to minimize the distance between a predicted solution and the solutions in the training set. Since a mixture of local and global solutions are used in training, the learned neural network would make a prediction that is the average of the two solutions. This predicted solution could very well end up being at the "wrong" place, as shown in Fig.~\ref{fig:regression}, where it would lead to the local solution rather than the global one. Note that, we could potentially preprocess the training data, but that is likely to be cumbersome and removes some of the appeal of using machine learning.

\captionsetup[figure]{font=small,skip=2pt}

\begin{figure}[ht]
\centering
\begin{subfigure}[b]{0.5\textwidth}
 \centering
 \includegraphics[height=5cm, width=6cm]{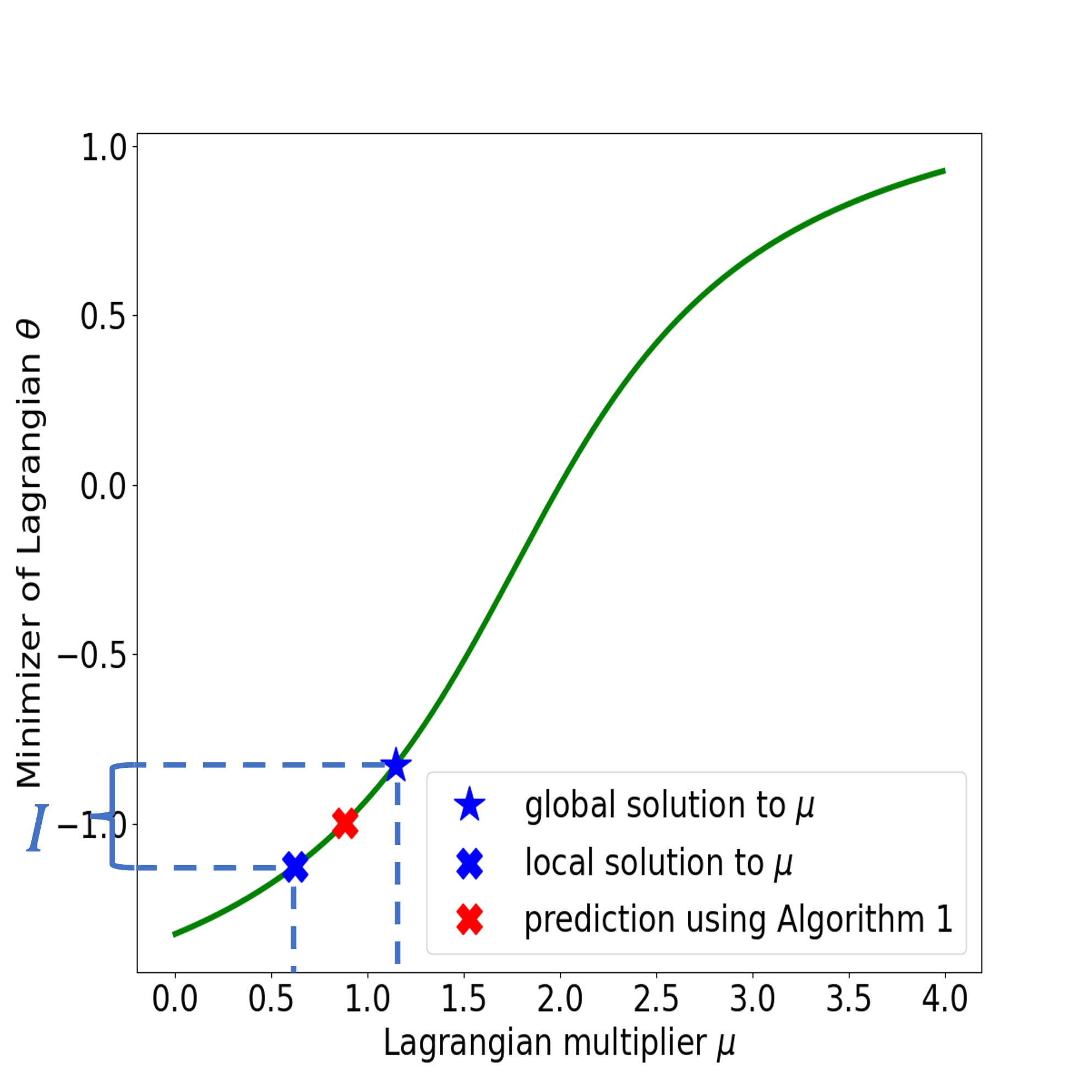}
 \caption{Learn the mapping from the Lagrangian multipliers to the solutions of the partial Lagrangian. When the training data set is a mix of local and global solutions, the predicted multiplier and the associated solution of the partial Lagrangian are marked as red. The set $I$ corresponds to the set of solutions of the partial Lagrangian when the multiplier varies between the locally and the globally optimal values.}
 \label{fig:algo1_1}
\end{subfigure}
\hfill
\begin{subfigure}[b]{0.5\textwidth}
 \centering
 \includegraphics[height=5cm, width=6cm]{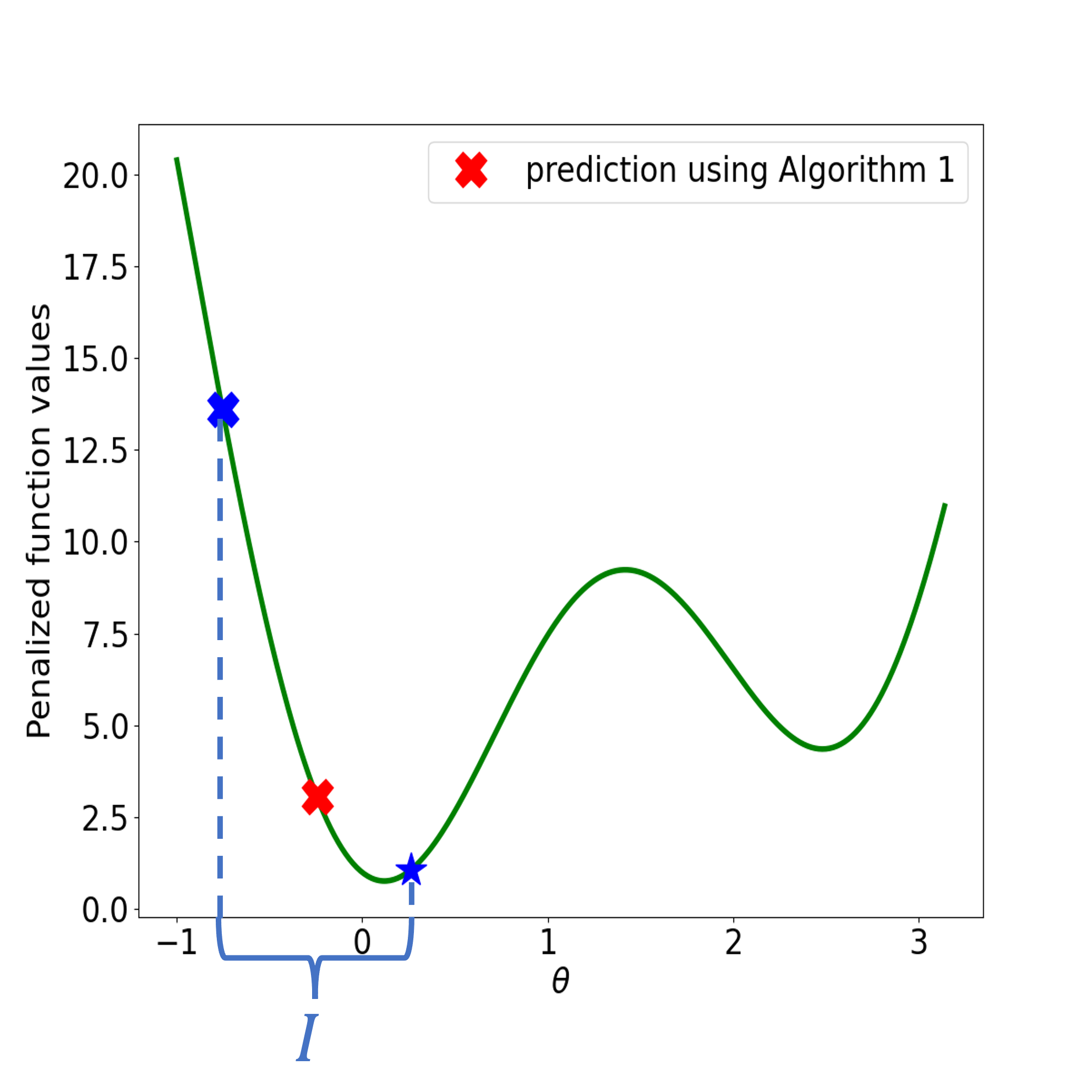}
\caption{
The learned warm start point using Algorithm 1 is marked as red, which is close to the global minimum of the ACOPF problem. If this predicted warm start is used, the solver would converge to the global solution.
In fact, the set $I$ is in the basin of  attraction of the global minimum.}
\label{fig:algo1_2}
\end{subfigure}
\caption{Use Algorithm 1 to learn a warm start point for the ACOPF solver.}
\label{fig:learning}
 \vspace{-0.3cm}
\end{figure}

Now we show what would happen when Algorithm 1 is used on this mixed data set.
For Algorithm 1, we first predict the multipliers from the load. Since the training set is a mix of local and global solutions, the predicted multiplier would be some point lying between the locally and globally optimal values. The predicted multiplier is plotted in Fig.~\ref{fig:algo1_1}.
Then we predict the solution of $\mathcal{L}_{\mu}$ from the predicted multiplier. 
For a given multiplier $\bar{\mu}$, the solution to $\mathcal{L}_{\mu}$ can be solved by writing out the optimality condition of \eqref{eqn:Lagrangian2bus}:
\begin{equation}\label{eq:optimality}
    (c^{\prime}+\bar{\mu}) g \sin(\bar{\theta})+(c^{\prime}-\bar{\mu}) b \cos(\bar{\theta})=0,
\end{equation}
where $c^{\prime}$ is a shorthand for $c^{\prime}(g-g\cos(\bar{\theta})+b\sin(\bar{\theta}))$ and is the gradient of the cost function.
By varying the multipliers, we can represent the mapping from the multipliers to the solutions of $\mathcal{L}_{\mu}$ as follows:
\begin{align}\label{eq:minimizer}
  \bar{\theta}=\tan^{-1}(\frac{\bar{\mu}-c^{\prime}}{\bar{\mu}+c^{\prime}}b/g).
\end{align}
The mapping function \eqref{eq:minimizer} is plotted in Fig.~\ref{fig:algo1_1}. 
The set of solutions of $\mathcal{L}_{\mu}$ that are mapped from 
the multipliers varying between the locally and globally optimal values is denoted by set $I$.
The predicted solution of $\mathcal{L}_{\mu}$ would lie in set $I$.
We also plot set $I$ in Fig.~\ref{fig:algo1_2}. We can see that every point in set $I$ is close to the global minimum of $\mathcal{L}_\rho$. More precisely, every point is in the basin of attraction of the global solution. 
This means if we use the predicted solution of $\mathcal{L}_{\mu}$ as a warm start, the solver would converge to the global minimum. 

In the next section, we test Algorithm 1 on IEEE benchmark systems, and show the intuition developed in this section is true for much larger and more complex problems. The simulation results show that Algorithm 1 is able to obtain globally optimal solutions even when the data set is only comprised of strictly local solutions.

\section{Simulation Results} \label{sec:results}
In this section, we 
demonstrate the simulation results of using Algorithm 1 to predict solutions to the ACOPF problem. We test our algorithm on IEEE networks with 22, 39 and 118 buses. 
The full specifications for these networks can be found in \cite{bukhsh2013local} and \cite{nguyen2014appearance}.
The popular solver IPOPT \cite{wachter2006implementation} is used to generate training samples for each network.
For the 22 and 118-bus networks, there exist more than one solution for a given load input.
We construct data sets comprised of different proportions of local solutions, and show the performance of Algorithm 1. For a comparison baseline, we use the method in \cite{baker2020emulating}, where a deep neural network is trained to learn the mapping from load to optimal generation values by minimizing the loss between the learned and ground-truth values. Then power flow equations are solved to recover and ensure feasibility of the overall ACOPF solutions.
We would see that our method can obtain globally optimal solutions even when the training data only contains local solutions.
For the 39-bus network, we would show that using the warm starts learned by our algorithm can speed up the computation time of solving ACOPF problems using IPOPT. 

We use fully-connected neural networks with 2 hidden layers for both Algorithm 1 and the baseline method. For the baseline method,
the activation function of the neural network is sigmoid for all layers. For Algorithm 1, we use ReLU as activation function except for the output layer, where linear activation function is used. All neural network models are implemented using Tensorflow software library.


\subsection{22 bus}
\captionsetup[figure]{font=small,skip=2pt}
\begin{figure}[t]
\centering
\includegraphics[height=5cm, width=6cm]{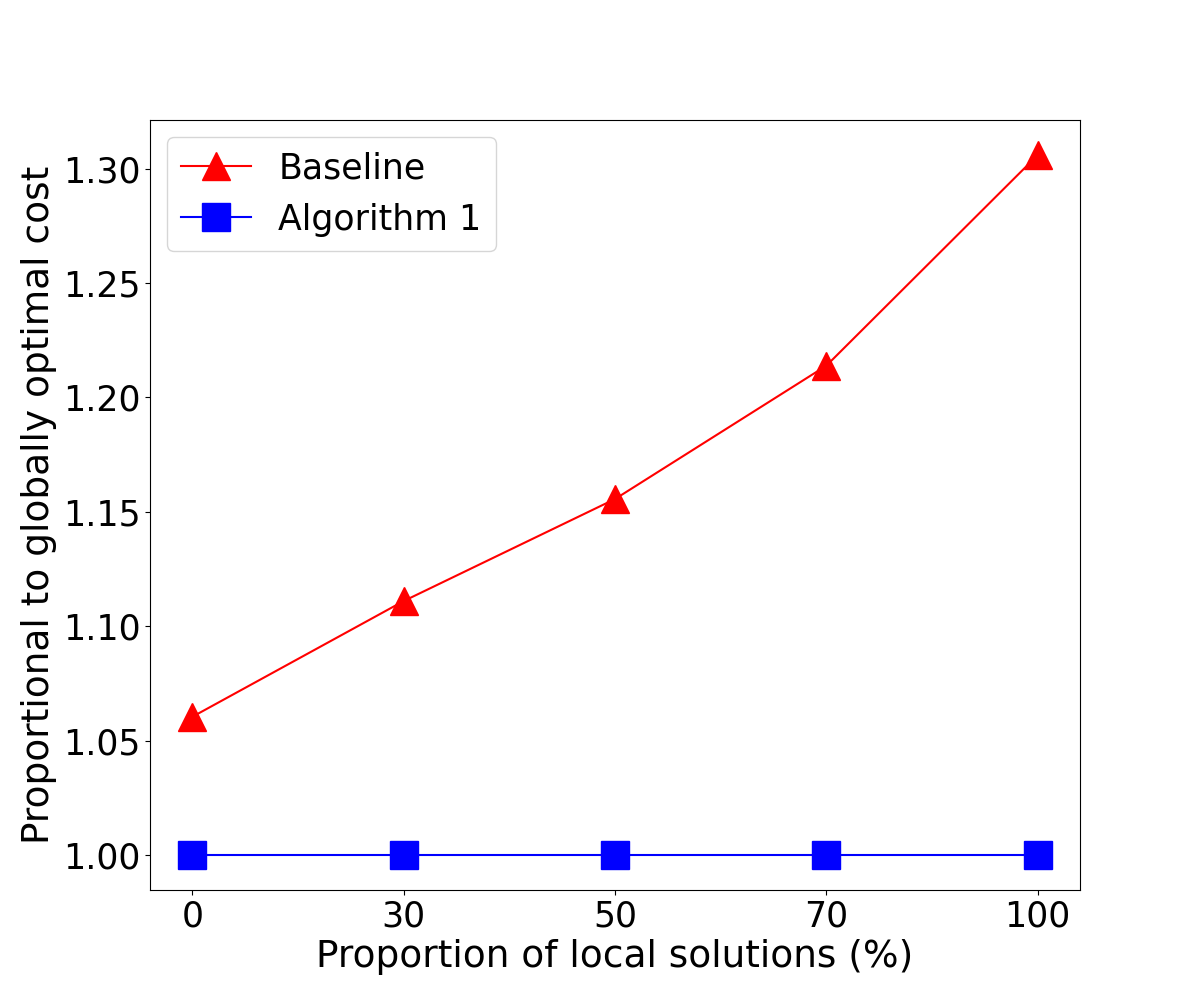}
\caption{Generation costs of the obtained solutions using Algorithm 1 on different training sets for 22-bus network. The predicted costs using the baseline method are also reported for comparison. All the generation costs are represented proportional to the globally optimal cost. We can see that Algorithm 1 is not sensitive to the quality of training data. It is able to obtain the global solution even when the training data is only consisted of local solutions.
\vspace{-0.7cm}}
\label{fig:22cost}
\end{figure}
In the 22-bus network, there exist two solutions for a given load. The cost
of the local solution is 30$\%$ higher than that of the global solution. We generate the training data by varying the load around the nominal value with $1\%$ variations using uniform distribution. For each given load, we solve the ACOPF problem using IPOPT to obtain both solutions. Then we
construct $5$ different training sets by adjusting the proportion of strictly local solutions in the data. There are 4000 training samples. We use $90\%$ of them for training and $10\%$ for testing. 
We compare the obtained solution quality using Algorithm 1 to that of the baseline method.

The obtained solutions using Algorithm 1 are always feasible, since the ACOPF problem is solved in the last step.  
In the baseline method, the power flow equations are solved to ensure feasibility of the solution. However, the active power generation is directly predicted using the neural network in the baseline and will not be affected by the power flow step. 
For simplicity, we assume the obtained solutions using the baseline method are feasible, and compare the predicted generation costs to the globally optimal cost.

The generation costs of the obtained solutions using both methods on different training sets are reported in Fig.~\ref{fig:22cost}, where the generation costs are represented proportional to the globally optimal cost. In Fig.~\ref{fig:22cost},
as the proportion of strictly local solutions in the training set increases, the predicted cost using the baseline method also increases, and is larger than the globally optimal cost on every training set. In contrast, Algorithm 1 is able to obtain the global solution, even when the training set is comprised only of strictly local solutions. This implies that Algorithm 1 is not sensitive to the quality of the training set, and local solutions can also be useful. 
This observation carries over to larger networks and we would test our algorithm on 118-bus network in Section \ref{sec:118}.

\subsection{39 bus}
\begin{figure}[t]
\centering
\includegraphics[height=5cm, width=6cm]{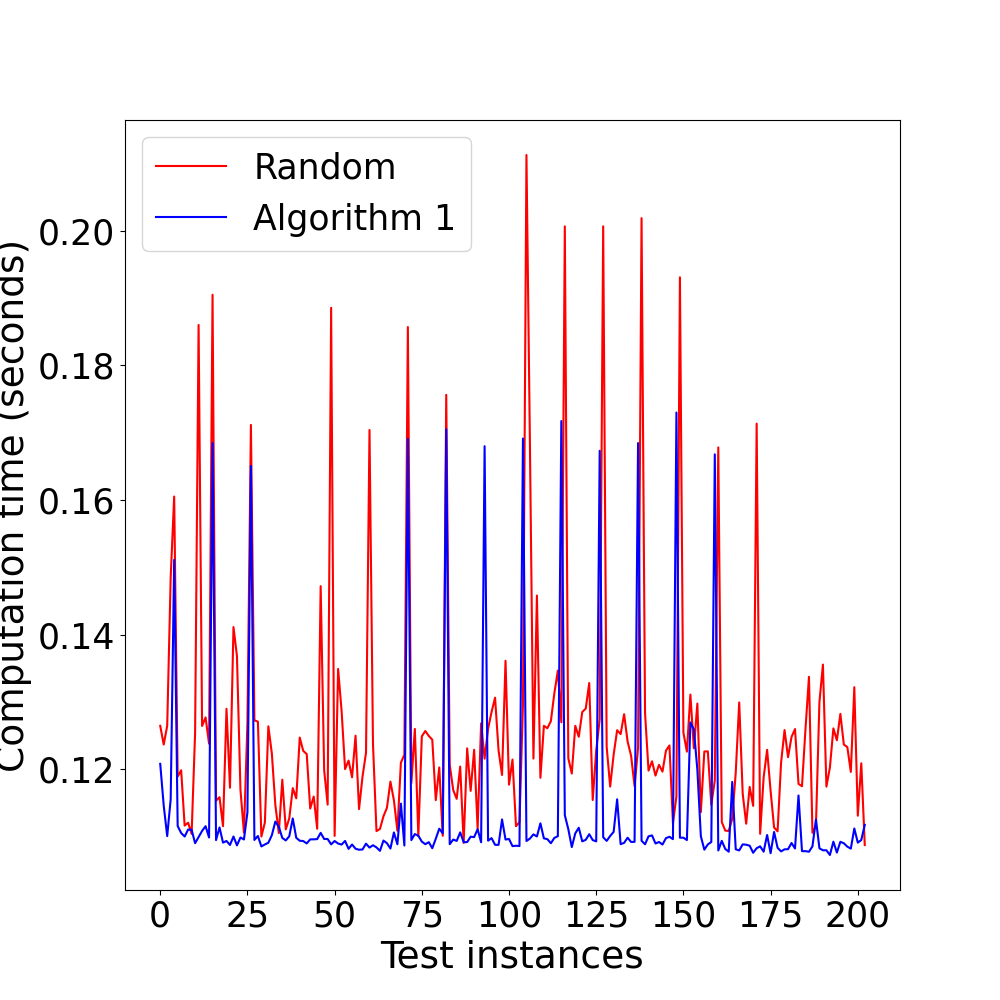}
\caption{Computation time of calling IPOPT to solve the ACOPF problem in 39-bus network with different initialization. The blue curve is the computation time when the warm starts learned by Algorithm 1 are used as initial points, which is lower than the random initialization (red curve) almost for every instance.
\vspace{-0.5cm}}
\label{fig:39time}
\end{figure}

For the 39-bus network that we use in this paper, there is only one solution. 
Since there is no difference in obtained solution quality when initialization is different, we use this network to demonstrate the computation time speed-up of calling IPOPT with the learned warm start points using Algorithm 1 as initialization.
In comparison, we also initialize IPOPT with randomly generated data points using Gaussian distribution. 
We evaluate the computation time on Macbook Pro with Intel Core i5 8259U CPU @ 2.30GHz. 
We call IPOPT with both initialization for $200$ instances and report the computation time for each instance in Fig.~\ref{fig:39time}. The computation time of using the learned warm starts given by Algorithm 1 is plotted as the blue line, which is faster than the random initialization (red) almost for every instance. In many cases, the learned warm start is much faster than random initialization. Note that the neural networks used in Algorithm 1 are feed-forward functions, and their evaluation time (sub-milliseconds) is negligible for the comparison in Fig.~\ref{fig:39time}.

To evaluate computational improvements, we also compute the average relative speedup of computation time 
between the two initialization methods.
Using the learned warm start point from Algorithm 1 as initialization can provide an on average $12\%$ speedup in computation time.
This is because
the minimizer of the Lagrangian would be close to the global solution of the ACOPF problem as we discussed in Section \ref{sec:geometry}. Therefore, using the predicted solution of the Lagrangian as the initial point enables IPOPT to find the solution in a shorter time.

\captionsetup[figure]{font=small,skip=2pt}

\subsection{118 bus}\label{sec:118}
For the 118-bus network, there exist three solutions for a given load. The worst cost is $39\%$ higher than the globally optimal cost.
We generate the training data by varying the load around the nominal value with $0.5\%$ variations using uniform distribution. For each given load, we solve the ACOPF problem using IPOPT to obtain the solution with the highest cost and the global solution.
Then we construct $3$ different training sets. Each of these training sets has half or more than half local solutions with the highest cost.
There are $1900$ training samples. We use $90\%$ of them for training and $10\%$ for testing.
We compare the generation costs of the obtained solutions using Algorithm 1 to that predicted by the baseline method. All the generation costs are represented in proportional to the globally optimal cost.

\begin{figure}[t]
\centering
\includegraphics[height=5cm, width=6cm]{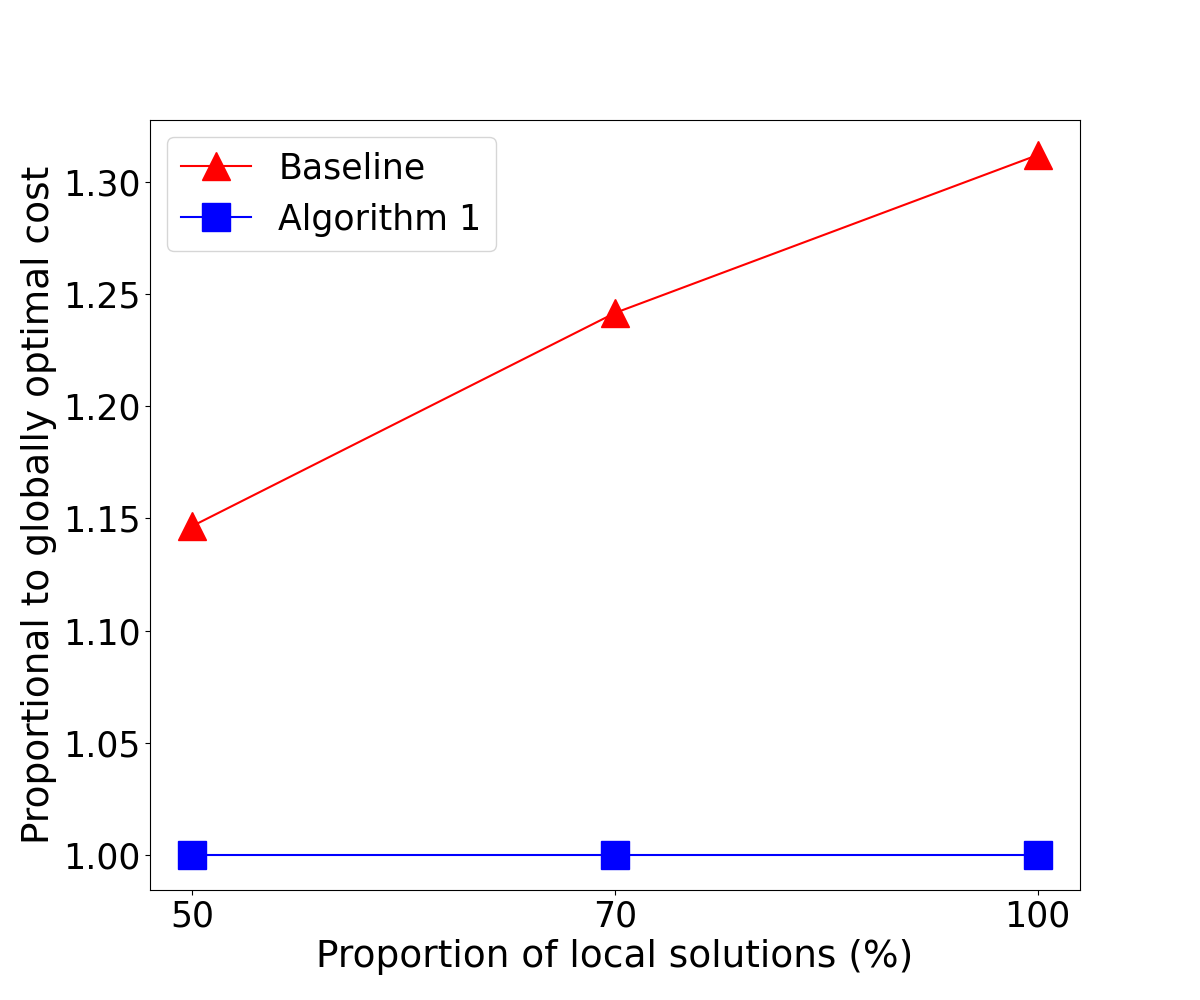}
\caption{Generation costs of the obtained solutions using Algorithm 1 and the baseline method on different training sets for 118-bus network. All the generation costs are represented proportional to the globally optimal cost. Algorithm 1 is  able to obtain the global solution even when the training data is only consisted of local solutions. 
\vspace{-0.5cm}}
\label{fig:118cost}
\end{figure}

The predicted generation costs using the two methods are plotted in Fig.~\ref{fig:118cost}. 
The predicted cost using the baseline method is larger than the globally optimal cost on every training set, and increases as the the quality of the training data declines (the proportion of local solutions increases). 
Since the neural network in the baseline method is trained by minimizing the loss between learned and ground-truth generation values, it is not a surprise that the predicted cost using the baseline method increases as the proportion of local solutions increases.
In contrast, Algorithm 1 is able to obtain the globally optimal cost regardless of the quality of the training data.
Even when the training data only contains the solutions with the highest cost, the predicted cost using Algorithm 1 is globally optimal.
This is because Algorithm 1 does not predict solutions to the ACOPF problem, but predicts solutions to the partial Lagrangian. 
As we discussed in Section \ref{sec:geometry}, 
the predicted solutions of the partial Lagrangian would be close to the global minimum of the ACOPF problem, and hence  could be good warm starts for the solver to reach the global solution.

\captionsetup[figure]{font=small,skip=2pt}

\section{Conclusion}
\label{sec:conclusion}
In this paper, we propose a partial Lagrangian-based learning approach to predict solutions of the ACOPF problem. First, we use a neural network to learn dual variables of the ACOPF problem. Then we use a second neural network to predict solutions of the partial Lagrangian from the predicted dual variables. 
Using the predicted solutions of the partial Lagrangian as warm starts, the ACOPF solver can reach more globally optimal solutions.
We illustrate the intuition behind our learning approach using a 2-bus network, which shows that the optimization landscape of the partial Lagrangian is much better than the original primal
problem. We validate the effectiveness of our algorithm
on standard 22-bus, 39-bus and 118-bus networks. 
The simulation results show that our algorithm is able to obtain the globally optimal cost even when the training data is only comprised of suboptimal solutions.

\bibliographystyle{IEEEtran}
\bibliography{mybib.bib}

\end{document}